\documentclass[showpacs,twocolumn,prl,aps]{revtex4}
\usepackage{graphicx}

\begin{document}

\title[Short title for running header]{Theory of the quasiparticle excitation in high T$_{c}$ cuprates: quasiparticle charge and nodal-antinodal dichotomy}
\author{Fan Yang$^{1}$ and Tao Li$^{2}$}

\affiliation{ $^{1}$Department of Physics, Beijing Institute of
Technology, Beijing 100081, P.R.China\\
 $^{2}$Department of Physics,
Renmin University of China, Beijing 100872, P.R.China}
\date{\today}

\begin{abstract}
A variational theory is proposed for the quasiparticle excitation in
high T$_{c}$ cuprates. The theory goes beyond the usual Gutzwiller
projected mean field state description by including the spin-charge
recombination effect in the RVB background. The spin-charge
recombination effect is found to qualitatively alter the behavior of
the quasiparticle charge as a function of doping and cause
considerable anisotropy in quasiparticle weight on the Fermi
surface.
\end{abstract}
\maketitle

\section{I. Introduction}
High temperature superconductors are doped Mott insulators. For such
a strongly correlated electron system, it is of fundamental
importance to know if and how does the Landau quasiparticle
excitations emerge and what is peculiar about their properties in
case of their existence. In recent decades, experiments, especially
the angle resolved photoemission(ARPES) and the scanning tunneling
microscopic spectrum(STM) measurements have elucidated in great
details the evolution of the quasiparticle properties of the
high-T$_{c}$ cupartes as a function of doping and
temperature\cite{Shen,Ding,Hanaguri,Kanigel}. However, a
comprehensive understanding of these experimental observations is
still in its infancy stage.

The existence of the quasiparticle excitation is now well
established in the superconducting state of the cuprates\cite{Shen}.
In the overdoped regime, the observations basically agrees with our
expectations for a conventional Landau Fermi liquid, where a large
Fermi surface enclosing a volume consistent with the Luttinger
theorem for a system with $1-x$ electron per unit cell is observed.
Well defined quasiparticle peaks are found all around the Fermi
surface too. However, below optimal doping, parts of the Fermi
surface around the antinodal region become less and less clear with
decreasing doping. This phenomena, which is dubbed as
nodal-antinodal dichotomy, is tangled with the issue of the
existence of two gap scales and small hole pocket Fermi
surface\cite{Tacon,Tanaka,Taillefer}. While in the half filing
limit, in which the cuprates develops antiferromagnetic long range
order, it is generally believed that quasiparticle excitation exists
in the form of spin polaron and form small hole pocket Fermi
surface. How is this small hole Fermi surface evolved into the large
electron Fermi surface in the overdoped regime is long unresolved
problem in the study of the high temperature superconductivity.

Another issue about the quasiparticle excitation in high Tc cuprates
is their electrodynamic response kernel, or more loosely, the
quasiparticle charge\cite{Lee}. As a result of the Mott physics, the
zero temperature superfluid density of the cuprates scales roughly
linearly with the density of the doped holes $x$, rather than the
result of $1-x$ when the strong correlation effect is neglected.
Take it literally, this would imply that in the ground state each
electron carry a charge of order $x$ rather than one. So naively one
would expect that the quasiparticle excitation above the ground
state should also carry a charge of order $x$. However, measurements
of the temperature dependence of the superfluid density indicates
that this is not the case. The almost doping independence of the
slope of the superfluid density curve as a function of temperature
indicates that the electrodynamic response kernel of the nodal
quasiparticle is almost a constant in the underdoped regime.

The total mobile charge density of a doped Mott insulator is given
by density of the doped holes. This requirement can be trivially
satisfied by a small hole Fermi surface enclosing an area of $x$, on
which each quasiparticle carries a charge of order one, as happens
in the antiferromagnetic ordered state. However, according to the
Landau Fermi liquid theory, a system with an electron density of
$1-x$ per unit cell and no symmetry breaking should always form a
large Fermi surface enclosing an area of $1-x$. So if we neglect the
possible anisotropy of the quasiparticle charge on the Fermi
surface, one would be naturally led to the conclusion that the
quasiparticle on the Fermi surface should each carry a charge of
order $x$. We are thus in a situation of dilemma. The system should
either violate the Luttinger theorem by possessing a small Fermi
surface of area $x$ in the absence of symmetry breaking, or, while
having a large Fermi surface that respects the Luttinger theorem,
have strong anisotropy on it. Both of these scenarios points to the
nontrivial nature of the quasiparticle properties in the cuprates.
It should be noted that suggestions have been made that in a system
with topological order, the Luttinger theorem should be modified to
accomodate the topological degeneracy\cite{Senthil,Paramekanti} and
a small Fermi surface in the absence of symmetry breaking is in
principle possible. However, up to date there is no evidence in
support of such a exotic order in the high T$_c$ cuprates.

The key to the physics of a doped Mott insulator is the local
constraint of no double occupancy between the electrons. The slave
Boson mean field theory, in which such a local constraint is treated
in an average manner, predicts a large Fermi surface consistent with
the Luttinger theorem\cite{Lee,Rantner}. The quasiparticle weight
and quasiparticle charge is however isotropic on the Fermi surface
and both show linear scaling with the hole density $x$. Attempts has
been made to go beyond the mean field theory by including the
fluctuation effect but the answer is still inconclusive\cite{Lee}.
The fluctuation around the mean field configuration, which takes the
form of gauge fluctuations, is notoriously hard to be analyzed as
there is no mass term to control their behavior.

The origin of the gauge degree of freedom in the slave Boson
formalism is just the no double occupancy constraint between the
electrons. To account for such a local constraint, Gutzwiller
projected mean field state of the form
$\mathrm{P}_{\mathrm{G}}\gamma^{\dagger}_{\mathrm{k},\sigma}|\mathrm{d-BCS}\rangle$
has been studied extensively to address the quasiparticle
problem\cite{Gros,Anderson,Randeria,Gros1,Yunoki,Nave,Shih,Sorella,Li}.
However, it is found that the Gutzwiller projected state inherits to
a large extent the properties of the mean field theory. For example,
it also predicts a large Fermi surface on which the quasiparticle
weight is uniform. Both the quasiparticle weight and the
quasiparticle charge still vanish in the half filling limit, albeit
with some powers different from the mean field theory predictions.

Despite these problems, both the slave Boson mean field theory and
the Gutzwiller projection scheme do capture one key feature of the
cuprates as a doped Mott insulator: the strong particle-hole
asymmetry in the vicinity of the Fermi energy, which is hard to
envisage in alternative scenarios where strong correlation effect is
ignored\cite{Rantner,Li,Anderson}. For this reason, we believe that
the slave Boson mean field theory and the Gutzwiller projected state
are good starting point for a more consistent theory.

An important consequence of the gauge fluctuation beyond the mean
field description is the so called spin-charge recombination
effect\cite{Lee,Wen}. The gauge fluctuation, which couples equally
to the spinon and the holon degree of freedom in the slave Boson
formalism, will induce mutual backflow effect between the two parts.
Such a backflow effect, which manifests itself as a kinematic effect
of the local constraint, is beyond the reach of the Gutzwiller
projection description and encourages the formation of spinon-holon
bound state, whose motion is less affected by the gauge fluctuation.

To account for such a spin-charge recombination effect, a RPA theory
has been proposed previously\cite{Lee2}. The theory introduced a
phenomenological attractive interaction between the spinon and the
holon to induce spinon-holon bound state which is interpreted as the
quasiparticle excitation of the system. Although the theory does
have the potential to explain certain features of the experiments,
it is not clear to what extent the predictions made by it are gauge
invariant. The prediction power of the theory is also limited by its
phenomenological nature and the sum rule for the electron spectral
function is in general violated. Alternatively, a dopon-spinon
formalism is introduced to account for the spin-charge recombination
effect in which the bare hole is treated as an elementary degree of
freedom\cite{Ran}. However, variational calculation based on this
formalism can only be proceeded in the half filling limit with a
small and finite number of doped holes.

In this paper, we extend the Gutzwiller projection description of
the quasiparticle by including the backflow effect between the
spinon and the holon at the wave function level. Our approach has
the advantage that its predictions are gauge invariant and the
electron spectral function calculated from it satisfy the relevant
sum rule. We found the quasiparticle charge after the backflow
effect correction approaches to a constant value in the half filling
limit, rather than vanishes as predicted by the mean field theory
and the Gutzwiller projection scheme. At the same time, the
quasiparticle weight is found to show more and more strong
anisotropy on the Fermi surface with decreasing doping and by fine
tuning of the Hamiltonian parameters one can indeed reproduce the
nodal-antinodal dichotomy phenomena in the cuprates.

This paper is organized as follows. In the next section, we review
the variational description of the quasiparticle excitation in the
Gutzwiller projection scheme and motivate our new variational wave
function by reformulating the Gutzwiller projected wave function in
terms of the slave Boson language in which the spin-charge
recombination effect can be easily included. In section III, we
introduce the numerical algorithm to do calculation on the wave
function we have proposed. In section IV, we present our numerical
results on both the doping dependence of the quasiparticle charge
and the anisotropy of the quasiparticle weight on the Fermi surface.
In section V, we present some discussion on the results. The
appendix contains some details of the derivations in the text.

\section{II. Variational quasiparticle wave function}
The model studied in this paper is the $t-J$ model,
\begin{equation}
\mathrm{H}=-\sum_{i,j,\sigma}t_{i,j}(\hat{c}^{\dagger}_{i,\sigma}\hat{c}_{j,\sigma}+h.c.)
+J\sum_{\langle i,j\rangle}(\vec{S}_{i}\cdot
\vec{S}_{j}-\frac{1}{4}n_{i}n_{j}),
\end{equation}
in which $\hat{c}_{i,\sigma}$ is the electron operator satisfying
the no double occupancy constraint
$\sum_{\sigma}\hat{c}^{\dagger}_{i,\sigma}\hat{c}_{i,\sigma}\leq 1$,
$t_{i,j}$ denotes the hopping integral between site $i$ and site
$j$. $\sum_{\langle i,j\rangle}$ denotes sum over nearest
neighboring sites. In this study, $t_{i,j}$ will be assumed to be
nonzero only between nearest neighboring, next nearest neighboring
and next next nearest neighboring sites. The corresponding hopping
integral will be denoted as $t$, $t^{'}$ and $t^{''}$.

The Gutzwiller projected BCS mean field state of the form
\begin{equation}
|\Psi\rangle=\mathrm{P}_{N_{e}}\mathrm{P}_{\mathrm{G}}|\mathrm{BCS}\rangle,
\end{equation}
is generally believed to be a good starting point for a variational
description of the ground state of the t-J model. Here
$\mathrm{P}_{\mathrm{G}}$ denotes the Gutzwiller projection into the
subspace satisfying the constraint
$\sum_{\sigma}\hat{c}^{\dagger}_{i,\sigma}\hat{c}_{i,\sigma}\leq 1$
and $\mathrm{P}_{N_{e}}$ denotes the projection into the subspace of
$N_{e}$ electrons. $|\mathrm{BCS}\rangle$ is the usual BCS mean
field ground state with d-wave pairing.

A variational description of the quasiparticle excitation on
$|\Psi\rangle$ can be constructed in the same spirit as
$|\Psi\rangle$ by Gutzwiller projection of the mean field excited
state. For example, the variational wave function for the
quasiparticle excitation of hole type has the form,
\begin{equation}
|\mathrm{k},\sigma\rangle=\mathrm{P}_{N_{e}-1}\mathrm{P}_{\mathrm{G}}\gamma^{\dagger}_{\mathrm{k},\sigma}|\mathrm{BCS}\rangle,
\end{equation}
in which
$\gamma^{\dagger}_{\mathrm{k},\sigma}=u_{\mathrm{k}}c^{\dagger}_{\mathrm{k},\sigma}+v_{\mathrm{k}}c_{-\mathrm{k},\bar{\sigma}}$.
This state has been studied by many authors in recent years. It is
shown that $|\mathrm{k},\sigma\rangle$ inherits to a large extent
the properties of the mean field excitation. In particular, the
quasiparticle weight is predicted to be isotropic on the underlying
Fermi surface. At the same time, both the quasiparticle weight and
the quasiparticle charge response kernel vanish in the half filling
limit, albeit with some powers different from that predicted by the
mean field theory.

To go beyond the Gutzwiller projected wave function and to make
connections with the effective field theory considerations, we
reformulate the Gutzwiller projection in the slave Boson language.
In the slave Boson formulation, the electron operator is expressed
in terms of the Fermionic spinon operator and the Bosonic holon
operator as
\begin{equation}
c_{i\sigma}=f_{i,\sigma}b^{\dagger}_{i}.
\end{equation}
The no double occupancy constraint now takes the form of an equality
\begin{equation}
\sum_{\sigma}f^{\dagger}_{i,\sigma}f_{i,\sigma}+b^{\dagger}_{i}b_{i}=1.
\end{equation}

In the slave Boson language, the t-J model takes the form
\begin{eqnarray}
\mathrm{H}&=&-\sum_{i,j,\sigma}t_{i,j}(f^{\dagger}_{i,\sigma}f_{j,\sigma}b^{\dagger}_{j}b_{i}+h.c.)\nonumber\\
&+&\frac{J}{2}\sum_{\langle i,j
\rangle}(f^{\dagger}_{i,\sigma}f^{\dagger}_{j,\sigma'}f_{j,\sigma'}f_{i,\sigma}-f^{\dagger}_{i,\sigma}f_{i,\sigma}f^{\dagger}_{j,\sigma'}f_{j,\sigma'}).
\end{eqnarray}
In the mean field treatment, the ground state of the $t-J$ model is
given by the product of the BCS mean field state for the Fermionic
spinon and the Bose-Einstein condensate of the holon. The
variational ground state $|\Psi\rangle$ can be shown to be given by
the Gutzwiller projection of such a product state.
\begin{eqnarray}
|\Psi\rangle&=&\mathrm{P}_{N_{h}}\mathrm{P}_{s}|f-\mathrm{BCS}\rangle\otimes|b-\mathrm{BEC}\rangle,
\end{eqnarray}
in which $|f-\mathrm{BCS}\rangle$ is the BCS mean field state of the
spinon and $|b-\mathrm{BEC}\rangle$ is the Bose condensate of the
holon. For notational convenience, in the following we will
abbreviate $|f-\mathrm{BCS}\rangle\otimes|b-\mathrm{BEC}\rangle$ as
$|\mathrm{SBMF}\rangle$. Here $\mathrm{P}_{s}$ denotes the
projection into the subspace satisfying the constraint
$\sum_{\sigma}f^{\dagger}_{i,\sigma}f_{i,\sigma}+b^{\dagger}_{i}b_{i}=1$
and $\mathrm{P}_{N_{h}}$ denotes the projection into the subspace
with $N_{h}$ doped holes. The BCS state for spinon is of the form
\begin{equation}
|f-\mathrm{BCS}\rangle=\left(
\sum_{i,j}a(i-j)f^{\dagger}_{i,\uparrow}f^{\dagger}_{j,\downarrow}\right)^{(N-N_{h})/2}|0\rangle,
\end{equation}
in which
$a(i-j)=\sum_{\mathrm{k}}\frac{v_{\mathrm{k}}}{u_{\mathrm{k}}}e^{i\mathrm{k}\cdot(\mathrm{R}_{i}-\mathrm{R}_{j})}$.
Here
$\frac{v_{\mathrm{k}}}{u_{\mathrm{k}}}=\frac{\Delta_{\mathrm{k}}}{\xi_{\mathrm{k}}+\sqrt{\xi^{2}_{\mathrm{k}}+\Delta^{2}_{\mathrm{k}}}}$
,
$\xi_{\mathrm{k}}=\frac{1}{N}\sum_{i,j}t^{v}_{i,j}e^{i\mathrm{k}\cdot(\mathrm{R}_{i}-\mathrm{R}_{j})}-\mu^{v}$
,
$\Delta_{\mathrm{k}}=\frac{1}{N}\sum_{i,j}\Delta^{v}_{i,j}e^{i\mathrm{k}\cdot(\mathrm{R}_{i}-\mathrm{R}_{j})}$.
$t^{v}_{i,j}$ and $\Delta^{v}_{i,j}$ are hopping and pairing
parameters determining the mean field ground state and are treated
as variational parameters to be optimized from the variational
energy, $\mu^{v}$ is also a variational parameter and not to be
mistaken as the real chemical potential. In this work, $t^{v}_{i,j}$
will be assumed to have the same range as the real hopping integral
$t_{i,j}$ and $\Delta^{v}_{i,j}$ will be assumed to take the
standard d-wave form.

In the slave Boson language, an electron becomes a composite object.
To create a hole in the system, one should generate a Bogliubov
quasiparticle in the BCS mean field ground state of the spinon and
at the same time add a holon to the system. The added holon can
either enter the Bose condensate of the holon or stay out of it. The
first choice for the holon leads to the coherent quasiparticle peak
in the electron spectral function. The variational wave function
$|\mathrm{k},\sigma\rangle$ for the quasiparticle is just given by
the Gutzwiller projection of this mean field state, \textit{i.e.},
\begin{equation}
|\mathrm{k},\sigma\rangle=\mathrm{P}_{N_{h}+1}\mathrm{P}_{s}\gamma^{\dagger}_{\mathrm{k},\sigma}b^{\dagger}_{\mathrm{q=0}}|\mathrm{SBMF}\rangle,
\end{equation}
here
$\gamma^{\dagger}_{\mathrm{k},\sigma}=u_{\mathrm{k}}f^{\dagger}_{\mathrm{k},\sigma}+v_{\mathrm{k}}f_{-\mathrm{k},\bar{\sigma}}$,

Up to this point, the slave Boson language seems to generate no new
result beyond the usual Gutzwiller projection scheme. To see the key
difference between the two schemes, we note that in the usual
Gutzwiller projection scheme, the commutator between the electron
operator and the Gutzwiller projection operator is nonzero,
$[\mathrm{P}_{\mathrm{G}},\hat{c}_{i,\sigma}]\neq0$, while in the
slave Boson language, the commutator between the electron operator
and the projection operator $\mathrm{P}_{s}$ is identically zero as
a result of the gauge invariance of the electron operator. Such a
property can be very useful when proving certain sum rules that will
be exemplified below. A simple application of this property leads
readily to the conclusion that the electron spectrum in the particle
side is totally coherent in the Gutzwiller projection scheme(as
holons are all condensed)\cite{Yunoki}.

One more advantage of the slave Boson language is that it provides a
bridge between the variational study and the effective field theory
considerations. For example, to describe the spin-charge
recombination effect argued in the effective field theory context,
we can introduce the following wave function
\begin{equation}
|\mathrm{k},\sigma\rangle_{\mathrm{scr}}=\mathrm{P}_{N_{h}+1}\mathrm{P}_{s}\sum_{\mathrm{q}}\phi_{\mathrm{q}}\gamma^{\dagger}_{\mathrm{k-q},\sigma}b^{\dagger}_{\mathrm{q}}|\mathrm{SBMF}\rangle,
\end{equation}
in which $\phi_{\mathrm{q}}$ can be interpreted as the wave function
for the relative motion between the spinon and holon. We note the
form of the wave function is quite general. For example, when
$\phi_{\mathrm{q}}=\delta_{\mathrm{q},0}$, the Gutzwiller projected
state $|\mathrm{k},\sigma\rangle$ is recovered, while when
$\phi_{\mathrm{q}}=v_{\mathrm{k-q}}$, the bare hole state
$|\mathrm{k},\sigma\rangle_{0}=c_{\mathrm{k},\sigma}\mathrm{P}_{N_{e}}\mathrm{P}_{\mathrm{G}}|\mathrm{BCS}\rangle$
is recovered(see Appendix A). In the following, we will take
$\phi_{\mathrm{q}}$ as variational parameters to be determined by
the optimization of energy. The spin-charge recombination effect
then manifests itself in the short ranged nature of the optimized
$\phi_{\mathrm{q}}$.

In the effective field theory description, the spin-charge
recombination effect is argued to be caused by the gauge
fluctuation, which acts to enforce the local constraint between the
spinon and the holon degree of freedom. In the Gutzwiller projection
scheme, such a local constraint is enforced a posteriori. By so
doing, the kinematic effect of the constraint, namely the backflow
effect between the spinon and the holon motion is totally missed. To
make this point more clearly, we note the kinetic part of the
Hamiltonian describes a correlated motion of the spinon and holon
\begin{equation}
\mathrm{H}_{\mathrm{K}}=-\sum_{i,j,\sigma}t_{i,j}(f^{\dagger}_{i,\sigma}f_{j,\sigma}b^{\dagger}_{j}b_{i}+h.c.),
\end{equation}
and the spinon current is exactly compensated by the backflow
current of the holon in each hopping steps. The main theme of the
present work is to elucidate the correction induced by the backflow
effect on the quasiparticle excitations.

In the slave Boson language, a hole-like quasiparticle is composed
of a spinon-holon pair. The backflow effect cause momentum transfer
between these two parts. In momentum space, the kinetic energy part
of the Hamiltonian is of the form
\begin{equation}
\mathrm{H}_{\mathrm{K}}=\frac{1}{N}\sum_{\mathrm{k,p,q},\sigma}\left[t(\mathrm{k+q-p})f^{\dagger}_{\mathrm{k+q},\sigma}f_{\mathrm{k},\sigma}b^{\dagger}_{\mathrm{p-q}}b_{\mathrm{p}}+h.c.\right],
\end{equation}
in which
$t(\mathrm{k})=\frac{1}{N}\sum_{i,j}t_{i,j}e^{i\mathrm{k}\cdot(\mathrm{R}_{i}-\mathrm{R}_{j})}$.
The scattering between the spinon and holon induced by this term
will in general lead to state state of the form
$\sum_{\mathrm{q}}\phi_{\mathrm{q}}\gamma^{\dagger}_{\mathrm{k-q},\sigma}b^{\dagger}_{\mathrm{q}}|\mathrm{SBMF}\rangle$
This is the reason that motivates the variational wave function
Eq.(10).

Two things should be noted here. First, in addition to causing
scattering between the existing spinon and holon pair, the backflow
effect can also generate extra pairs of spinon and holon from the
mean field ground state. This can be interpreted as a
renormalization of the ground state, which is not considered in the
present work. Such a renormalization effect can in principle be
taken into account in the variational wave function for the ground
state to arrive at a more consistent description of the
quasiparticle excitation. Second, it can be seen that the momentum
dependence of the backflow effect depends on the detailed form of
the hopping integral in the Hamiltonian. Thus some of the results
presented below are not generic, but depends on the Hamiltonian
parameters. This is especially the case for the nodal-antinodal
dichotomy phenomena. However, the quasiparticle charge around the
nodal point(from where the contribution to the in-plane transport is
the largest) is to a large extent not sensitive to the fine tuning
of the Hamiltonian parameters.

\section{III. Numerical algorithms}

The variational wave function proposed above is composed of $N$
Slater determinants and can be written as
\begin{equation}
|\mathrm{k},\sigma\rangle_{\mathrm{scr}}=\sum_{\mathrm{q}}\phi_{\mathrm{q}}|\mathrm{k,q},\sigma\rangle,
\end{equation}
in which
$|\mathrm{k,q},\sigma\rangle=\mathrm{P}_{N_{h}+1}\mathrm{P}_{s}\gamma^{\dagger}_{\mathrm{k-q},\sigma}b^{\dagger}_{\mathrm{q}}|\mathrm{SBMF}\rangle$
form a set of strongly correlated basis functions. Unlike the mean
field states
$\gamma^{\dagger}_{\mathrm{k-q},\sigma}b^{\dagger}_{\mathrm{q}}|\mathrm{SBMF}\rangle$,
$|\mathrm{k,q},\sigma\rangle$ are no longer orthogonal to each
other. Furthermore, it can be shown that not all
$|\mathrm{k,q},\sigma\rangle$ are linearly independent. The proof of
this point is left to the appendix.

In terms of this set of strongly correlated basis functions, the
variational energy for the quasiparticle is given by
\begin{equation}
\mathrm{E}_{\mathrm{k}}=\frac{\sum_{\mathrm{q,q'}}\phi^{*}_{\mathrm{q}}\phi_{\mathrm{q'}}\langle\mathrm{k,q},\sigma|\mathrm{H}_{t-J}|\mathrm{k,q'},\sigma\rangle}{\sum_{\mathrm{q,q'}}\phi^{*}_{\mathrm{q}}\phi_{\mathrm{q'}}\langle\mathrm{k,q},\sigma|\mathrm{k,q'},\sigma\rangle}.
\end{equation}
The minimization of this expression can be casted into a generalized
eigenvalue problem of the form\cite{Li1}
\begin{equation}
\sum_{\mathrm{q}}\mathrm{H}_{\mathrm{q},\mathrm{q'}}\phi_{\mathrm{q}}=\lambda\sum_{\mathrm{q}}\mathrm{O}_{\mathrm{q},\mathrm{q'}}\phi_{\mathrm{q}},
\end{equation}
in which
$\mathrm{H}_{\mathrm{q},\mathrm{q'}}=\langle\mathrm{k,q},\sigma|\mathrm{H}_{t-J}|\mathrm{k,q'},\sigma\rangle$
and
$\mathrm{O}_{\mathrm{q},\mathrm{q'}}=\langle\mathrm{k,q},\sigma|\mathrm{k,q'},\sigma\rangle$.
The optimized variational energy is given by the lowest eigenvalue
$\lambda_{min}$ of the generalized eigenvalue problem. The optimized
variational parameters $\phi_{\mathrm{q}}$ is given by the
eigenvectors corresponding to $\lambda_{min}$.

More generally, the calculation can be interpreted as
diagonalization of the Hamiltonian in the set of the strongly
correlated basis functions $|\mathrm{k,q},\sigma\rangle$. Thus if we
assume approximate completeness of the basis, we can even calculate
the full electron spectral function $A(\mathrm{k},\omega)$ as well.
The expression for $A(\mathrm{k},\omega)$ in this approximation is
given by
\begin{equation}
A(\mathrm{k},\omega)=\sum_{\mathrm{n,q,q'}}|\phi^{\mathrm{n}*}_{\mathrm{q}}\mathrm{O}_{\mathrm{q,q'}}\phi^{\mathrm{0}}_{\mathrm{q}}|^{2}\delta(\omega-(\lambda_{n}-\mathrm{E}_{g})),
\end{equation}
in which $\phi^{n}_{\mathrm{q}}$ denotes the normalized eigenvector
of Eq.(15) with eigenvalue $\lambda_{n}$. $\phi^{0}_{\mathrm{q}}$
denotes the vector corresponding to a bare hole on the RVB
background and is given by $\phi^{0}_{\mathrm{q}}=v_{\mathrm{k-q}}$,
$\mathrm{E}_{g}$ denotes the variational ground state energy. The
derivation of Eq.(16) is given in the appendix, in which it is also
shown that the spectral function so calculated satisfies the sum
rule of the form $\int d\omega A(\mathrm{k},\omega)=n_{\mathrm{k}}$.
The existence of such a sum rule partially justifies the approximate
completeness of the basis functions $|\mathrm{k,q},\sigma\rangle$.

The most time consuming part of the present calculation is the
determination of the Hamiltonian matrix elements
$\mathrm{H}_{\mathrm{q,q'}}$ and the overlap matrix elements
$\mathrm{O}_{\mathrm{q,q'}}$. In Ref \cite{Li1}, a highly efficient
reweighting technique to reach this goal is proposed based on the
mutual similarity of the basis functions. The algorithm has been
discussed in details in \cite{Li1}. Here we will only give a brief
overview of it.

To do variational Monte Carlo simulation, we expand the basis
function in a local basis $|R_{i}\rangle$ as
\begin{equation}
|\mathrm{k,q},\sigma\rangle=\sum_{i}\psi_{\mathrm{q}}(R_{i})|R_{i}\rangle.
\end{equation}
In our calculation, we have made a particle-hole transformation on
the down spin electron so that the wave function
$\psi_{\mathrm{q}}(R_{i})$ takes the form of a product of a Slater
determinant from the spinon part and a plane wave from the holon
part.

To calculate the overlap matrix elements, one can simulate the
following expression by Monte Carlo method
\begin{equation}
\frac{\mathrm{O}_{\mathrm{q,q'}}}{\mathrm{O}_{\mathrm{q,q}}}=\frac{\sum_{i}|\psi_{\mathrm{q}}(R_{i})|^{2}\frac{\psi_{\mathrm{q'}}(R_{i})}{\psi_{\mathrm{q}}(R_{i})}}{\sum_{i}|\psi_{\mathrm{q}}(R_{i})|^{2}}.
\end{equation}
However, there are of order $N^{2}$ such terms to be calculated and
a direct calculation is very time consuming. In \cite{Li1}, it is
shown that a more efficient and statistically more stable way to
calculate the overlap matrix elements is to simulate the following
expression
\begin{equation}
\frac{\mathrm{O}_{\mathrm{q,q'}}}{\sum_{\mathrm{q}}\mathrm{O}_{\mathrm{q,q}}}=\frac{\sum_{i}W(R_{i})\frac{\psi^{*}_{\mathrm{q}}(R_{i})\psi_{\mathrm{q'}}(R_{i})}{W(R_{i})}}{\sum_{i}W(R_{i})},
\end{equation}
in which $W(R_{i})=\sum_{\mathrm{q}}|\psi_{\mathrm{q}}(R_{i})|^{2}$.
The most important advantage of Eq.(19) over Eq.(18) is that the
simulation of the all $N^{2}$ matrix elements
$\mathrm{O}_{\mathrm{q,q'}}$ can now be done in a single run of the
Monte Carlo procedure. Another advantage is that the statistical
error of Eq.(18) caused by the nodes in $\psi_{\mathrm{q}}(R_{i})$
is now reduced.

To simulate the expression Eq.(19), we choose an arbitrary basis
function $|\mathrm{k,q_{0}},\sigma\rangle$ as a reference state,
then
\begin{equation}
W(R_{i})=|\psi_{\mathrm{q}_{0}}(R_{i})|^{2}\sum_{\mathrm{q}}\left|\frac{\psi_{\mathrm{q}}(R_{i})}{\psi_{\mathrm{q}_{0}}(R_{i})}\right|^{2},
\end{equation}
and
\begin{equation}
\frac{\psi^{*}_{\mathrm{q}}(R_{i})\psi_{\mathrm{q'}}(R_{i})}{W(R_{i})}=\frac{\left(\frac{\psi_{\mathrm{q}}(R_{i})}{\psi_{\mathrm{q}_{0}(R_{i})}}\right)^{*}\frac{\psi_{\mathrm{q'}}(R_{i})}{\psi_{\mathrm{q}_{0}(R_{i})}}}{\sum_{\mathrm{q}}\left|\frac{\psi_{\mathrm{q}}(R_{i})}{\psi_{\mathrm{q}_{0}}(R_{i})}\right|^{2}}.
\end{equation}

The Hamiltonian matrix elements can be simulated in a similar
manner. As Eq.(19), we have
\begin{equation}
\frac{\mathrm{H}_{\mathrm{q,q'}}}{\sum_{\mathrm{q}}\mathrm{O}_{\mathrm{q,q}}}=\frac{\sum_{i}W(R_{i})\frac{\psi^{*}_{\mathrm{q}}(R_{i})\times
\mathrm{H}\psi_{\mathrm{q'}}(R_{i})}{W(R_{i})}}{\sum_{i}W(R_{i})},
\end{equation}
in which $\mathrm{H}\psi_{\mathrm{q}}(R_{i})=\sum_{j}\langle
R_{i}|\mathrm{H}_{t-J}|R_{j}\rangle\psi_{\mathrm{q}}(R_{j})$, and
\begin{equation}
\frac{\psi^{*}_{\mathrm{q}}(R_{i})\times
\mathrm{H}\psi_{\mathrm{q'}}(R_{i})}{W(R_{i})}=\frac{\left(\frac{\psi_{\mathrm{q}}(R_{i})}{\psi_{\mathrm{q}_{0}(R_{i})}}\right)^{*}\frac{\mathrm{H}\psi_{\mathrm{q'}}(R_{i})}{\psi_{\mathrm{q}_{0}(R_{i})}}}{\sum_{\mathrm{q}}\left|\frac{\psi_{\mathrm{q}}(R_{i})}{\psi_{\mathrm{q}_{0}}(R_{i})}\right|^{2}}
\end{equation}

Thus to simulate the overlap matrix elements and the Hamiltonian
matrix elements, we only need to calculate the ratios
$\frac{\psi_{\mathrm{q}}(R_{i})}{\psi_{\mathrm{q}_{0}}(R_{i})}$ and
$\frac{\mathrm{H}\psi_{\mathrm{q}}(R_{i})}{\psi_{\mathrm{q}_{0}}(R_{i})}$in
each Monte Carlo steps. This can be easily done with the inverse
update trick as $|\mathrm{k,q},\sigma\rangle$ and
$|\mathrm{k,q_{0}},\sigma\rangle$ differs with each other by at most
a pair of spinon and holon states. In addition, these two set of
ratios form two vectors which make their calculation highly
parallelized.

The whole computational procedure can then be summarized as follows.
First, we determine the variational parameters in the ground state
by optimizing the ground state energy. Then we calculate the overlap
matrix and the Hamiltonian matrix with the reweighting technique.
With these matrixes, we solve the generalized eigenvalue problem.
The eigenvector corresponding to the lowest eigenvalue is the wanted
quasiparticle excitation in this scheme. From this wave function we
can then calculate the quasiparticle properties such as its weight,
charge, and dispersion relation. We can also use the basis as a
pseudo-complete one to approximate the electron spectral function to
gain an understanding of the incoherent part of the spectral
function.

\section{IV. Numerical results}
Our calculation is done on a $\sqrt{N}\times\sqrt{N}$ square lattice
with periodic-antiperiodic boundary condition. The Hamiltonian
parameters are chosen as follows. In our discussion of the nodal
quasiparticle properties, which are insensitive to the fine tuning
of the Hamiltonian parameters, we assume $t^{''}=0$ and set
$\frac{J}{t}=\frac{1}{3}$, $\frac{t'}{t}=-0.25$. However, when
discussing the antinodal quasiparticle properties, which are
sensitive to the value of $t^{''}$, we will present results for both
$t^{''}=0$ and $t^{''}=0.2t$(which is more realistic for the
cuprates). The doping concentration studied in this work ranges from
$2\%$ to $22\%$.

\subsection{A. The quasiparticle peak and the electron spectral function}
To illustrate the spin-charge recombination effect on the
quasiparticle properties, in Fig.1 we plot the eigenvalues of
Eq.(15) in ascending order for a system with $18\times18$ sites and
36 doped holes. The momentum of the quasiparticle is chosen along
the nodal direction and just below the Fermi surface. The
eigenvalues split into an isolated pole and a continuum with a
finite gap between them. The existence of the gap implies the
formation of spinon-holon bound state. The emergence of the bound
state can also be seen directly from the Fourier transform of
$\phi^{1}_{\mathrm{q}}$, which decreases exponentially with the
separation between the spinon and the holon.
\begin{figure}[h!]
\includegraphics[width=8cm,angle=0]{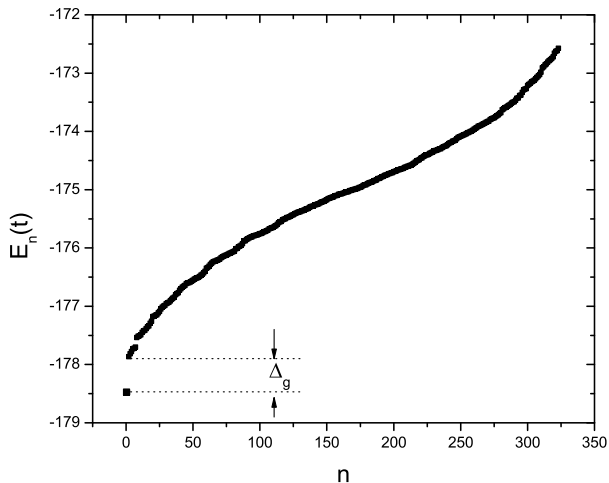}
\caption{The eigenvalues of Eq.(15) in ascending order for a system
with $18\times18$ sites and 36 doped holes. The momentum of the
quasiparticle is chosen along the nodal direction and slightly below
the Fermi surface. $t^{''}=0$ is assumed in the calculation.}
\label{fig1}
\end{figure}

The quasiparticle peak in the electron spectral function is just
contributed by this spinon-holon bound state. An approximate
electronic spectral function at the given momentum is shown in
Fig.2, showing clearly the emergence of the quasiparticle peak out
of incoherent background.

\begin{figure}[h!]
\includegraphics[width=8cm,angle=0]{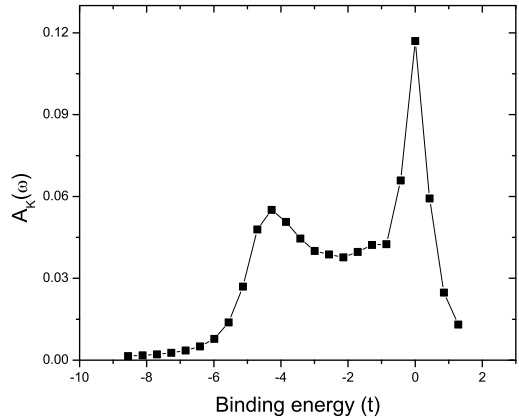}
\caption{The approximate electron spectral function calculated from
Eq.(16). The parameters used here are the same as those used in
Fig.1. The $\delta$-function peaks in Eq.(16) are broadened with a
width of $J$ in the calculation.} \label{fig2}
\end{figure}

\subsection{B. Quasiparticle charge}
The quasiparticle charge can be defined through the current carried
by it. In general, the current carried by a quasiparticle can be
written in the form $\vec{j}=q\vec{v}_{\mathrm{k}}$, where
$\vec{v}_{\mathrm{k}}=\nabla_{\mathrm{k}}\mathrm{E}_{\mathrm{k}}$ is
the group velocity of the quasiparticle calculated from its
dispersion relation and $q$ is the charge of the quasiparticle. As
the nodal quasiparticle has the largest velocity in the Brillouin
zone and thus dominates the in-plane electromagnetic response of the
system, our discussion of the quasiparticle charge will be
restricted to the nodal quasiparticles. As the velocity of the nodal
quasiparticle is almost independent of the hole concentration in the
high T$_{c}$ cuprates, we can use the current carried by the
quasiparticle as a measure of its charge.

\begin{figure}[h!]
\includegraphics[width=9cm,angle=0]{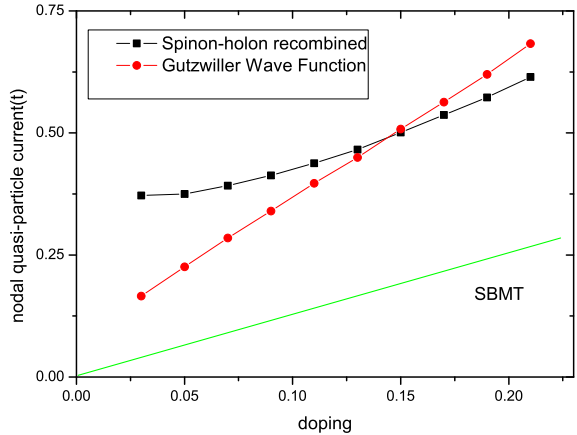}
\caption{The quasiparticle charge of the nodal quasiparticle as a
function of the hole concentration. The calculation is done on a
$14\times14$ lattice. $t^{''}=0$ is assumed in the calculation.}
\label{fig3}
\end{figure}

Since the property of the nodal quasiparticle is insensitive to the
fine tuning of the Hamiltonian parameters, we will assume $t^{''}=0$
in the following calculations. The electromagnetic current operator
of the t-J model in this assumption is given by
\begin{eqnarray}
j_{x}&=&it\sum_{i,\sigma}(c^{\dagger}_{i,\sigma}c_{i+x,\sigma}-h.c.)\nonumber\\
&+&it'\sum_{i,\sigma}(c^{\dagger}_{i,\sigma}c_{i+x+y,\sigma}-h.c.)\nonumber\\
&+&it'\sum_{i,\sigma}(c^{\dagger}_{i,\sigma}c_{i+x-y,\sigma}-h.c.),
\end{eqnarray}
with $j_{y}$ given by a similar expression. In the slave Boson mean
field theory, the current can be easily found to be given by
$\vec{j}=x\nabla_{\mathrm{k}}\epsilon_{\mathrm{k}}$, where
$\epsilon_{\mathrm{k}}$ is the mean field dispersion of the
quasiparticle. Thus in the slave Boson mean field theory, the
quasiparticle charge is proportional to the hole density $x$.

The quasiparticle current calculated from our variational wave
function is shown in Fig.3 in which the result is compared with the
predictions of the slave Boson mean field theory and the Gutzwiller
projected variational wave function $|\mathrm{k},\sigma\rangle$.
Unlike prediction of the mean field theory and the Gutzwiller
projection scheme, the quasiparticle charge calculated from our wave
function approaches to a finite value in the half filling limit. A
non-vanishing quasiparticle charge, which is consistent with
experimental observation in underdoped cuprates, constitutes the
main achievement of the present theory.

\subsection{C. The quasiparticle weight and the nodal-antinodal dichotomy}
In the slave Boson mean field theory, the quasiparticle weight is
given by $Z_{\mathrm{k}}=xv^{2}_{\mathrm{k}}$. Thus,
$Z_{\mathrm{k}}$ is isotropic on the underlying Fermi surface and
increases with the excitation energy below the Fermi surface. Apart
from some detailed difference in the doping dependence, the mean
field predictions on the quasiparticle weight are to a large extent
inherited by the Gutzwiller projected wave function
$|\mathrm{k},\sigma\rangle$. However, the increase of the
quasiparticle weight with excitation energy is obviously at odds
with the experimental observations and our physical intuitions. At
the same time, measurements have detected large anisotropy in the
quasiparticle weight on the underlying Fermi surface in the
underdoped cuprates. These two points constitute the main problems
with the Gutzwiller projected wave function
$|\mathrm{k},\sigma\rangle$.

To see if the backflow effect can cure these problems of
$|\mathrm{k},\sigma\rangle$, we have calculated the quasiparticle
weight from $|\mathrm{k},\sigma\rangle_{\mathrm{scr}}$ as a function
of momentum. The quasiparticle weight in the present theory reads
\begin{equation}
Z_{\mathrm{k}}=\frac{\left|_{0}\langle\mathrm{k},\sigma|\mathrm{k},\sigma\rangle_{\mathrm{scr}}\right|^{2}}{\langle\Psi|\Psi\rangle}.
\end{equation}

\begin{figure*}
\includegraphics[width=16cm,angle=0]{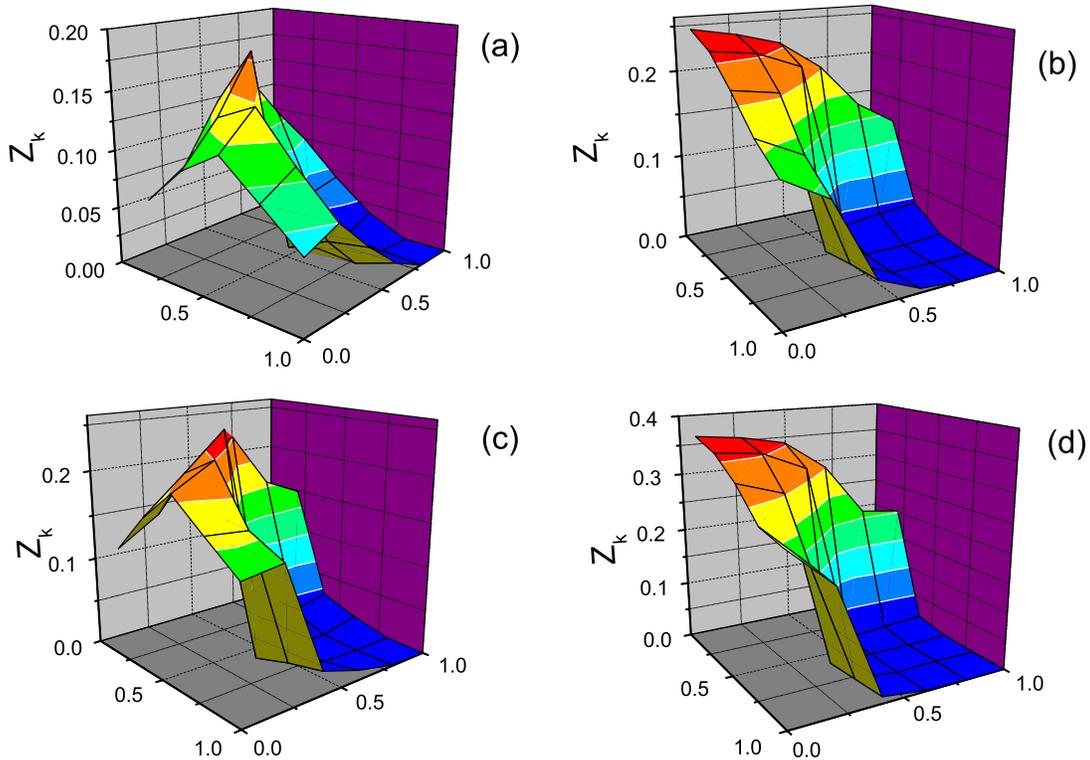}
\caption{The quasiparticle weight in the first quadrant of the
Brillouin zone for $t^{''}=0.2t$ calculated from the spin-charge
recombined wave function((a) and (c)) and the Gutzwiller projected
wave function((b) and (d)) at $x=6\%$((a) and (b)) and $x=14\%$((c)
and (d)). The calculation is done on a $10\times10$ lattice.}
\label{fig5}
\end{figure*}

As the properties of the off-nodal quasiparticles are sensitive to
the Hamiltonian parameters, we will present results calculated for
both the $t^{''}=0.2t$ and $t^{''}=0$ case. The results for
$t^{''}=0.2t$ is shown in Fig.5. Unlike the Gutzwiller projected
wave function, our variational wave function predicts a
quasiparticle weight that peaks on the Fermi surface. At the same
time, the quasiparticle weight is anisotropic on the Fermi surface.
The anisotropy is found to increases with decreasing doping and at
$x=6\%$ the quasiparticle weight in the nodal region is almost three
times larger than that in the antinodal region. These predictions of
our theory resemble closely the experimental observations.

However, the anisotropy of the quasiparticle weight on the Fermi
surface is not a generic property of our theory, but depends on fine
tuning of Hamiltonian parameters. In Fig.6 we show the quasiparticle
weight calculated with $t^{''}=0$. The anisotropy of the
quasiparticle weight on the Fermi surface is found to be much
smaller than that calculated with $t^{''}=0.2t$. Thus, in our theory
the nodal-antinodal dichotomy is not a generic consequence of the
spin-charge recombination effect, but depends on fine tuning of
Hamiltonian parameters. The same conclusion is also reached by the
variational calculation based on the dopon-spinon
formalism\cite{Ran}.

\begin{figure*}
\includegraphics[width=16cm,angle=0]{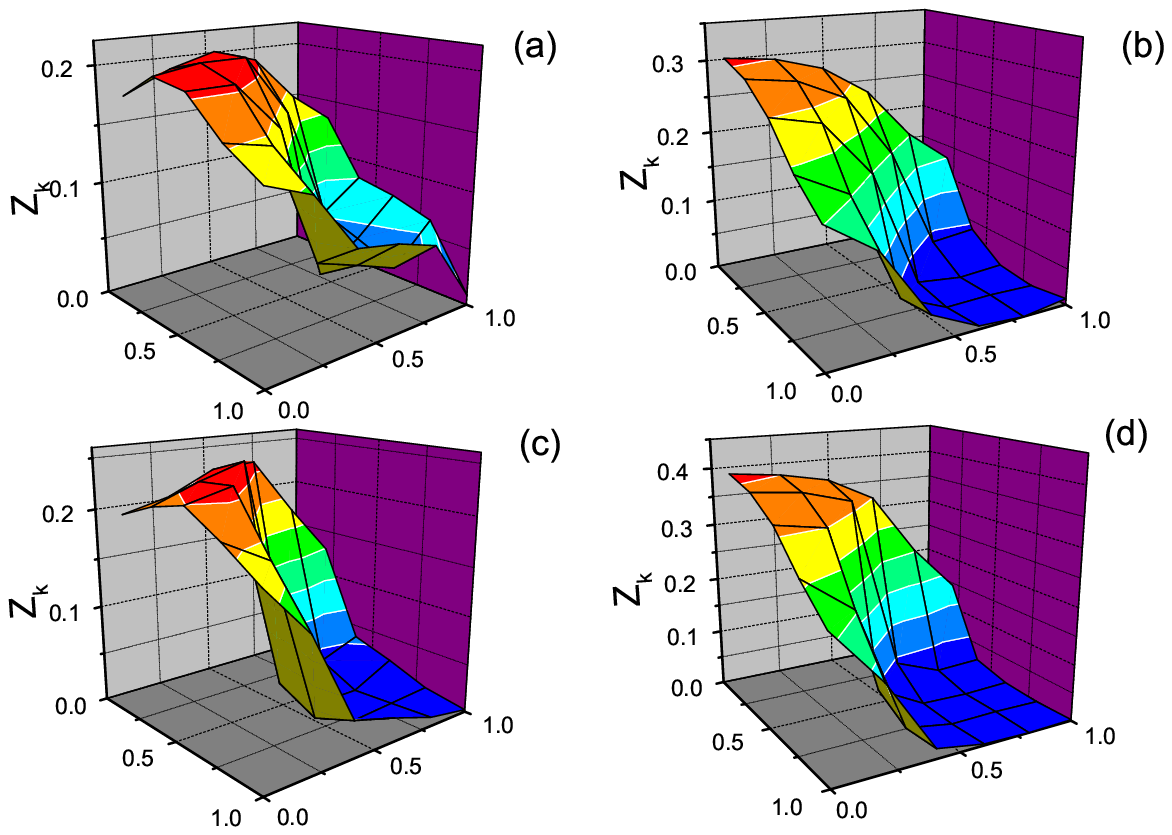}
\caption{The quasiparticle weight in the first quadrant of the
Brillouin zone for $t^{''}=0$ calculated from the spin-charge
recombined wave function((a) and (c)) and the Gutzwiller projected
wave function((b) and (d)) at $x=6\%$((a) and (b)) and $x=14\%$((c)
and (d)). The calculation is done on a $10\times10$ lattice.}
\label{fig6}
\end{figure*}

\section{V. Discussions}
In this work we have studied the consequence of the spin-charge
recombination effect on the quasiparticle properties of the
high-T$_{c}$ cuprates. The spin-charge recombination effect, which
can be interpreted as a backflow effect between the spinon and the
holon degree of freedoms, originates from the no double occupancy
constraint of the system. We find such a backflow effect will induce
spinon-holon bound state and will cause qualitative changes in the
quasiparticle properties.

In the Gutzwiller projected wave function, the no double occupancy
constraint is enforced by hand. However, such a posteriorly enforced
projection can not account for the full effect of the local
constraint. Especially, the kinematic effect of such a constraint on
the motion of the spinon and holon, namely the backflow effect, is
totally beyond the reach of such a description. In the effective
field theory context, the Gutzwiller projection amounts to
integration of the temporal component of the gauge fluctuation with
the spatial component of the gauge fluctuation totally untouched.
Such a unbalanced nature of the treatment of the gauge fluctuation
explains the qualitative similarity between the predictions made by
the mean field theory and the Gutzwiller projected wave function.

The most remarkable consequence of the backflow effect on the
quasiparticle properties is the modification of its electrodynamic
response kernel. In the mean field theory, in which the motion of
the spinon and holon is independent of each other, a quasiparticle
carries a charge of $x$ through the holon condensate. The vanishing
of the quasiparticle charge in the half filling limit predicted by
the mean field theory is inherited by the Gutzwiller projected wave
function. After the backflow effect correction, the spinon and the
holon form bound state. The holon dragged by the spinon in such a
composite object contributes a nonvanishing charge to the
quasiparticle even in the half filling limit. It should be
emphasized that this is generic consequence of the backflow effect
and does not depends on fine tuning of Hamiltonian parameters.

The backflow effect can also provide a potential mechanism for the
experimental observation of nodal-antinodal dichotomy and at the
same time result in a quasiparticle weight that decreases with
increasing excitation energy, as would be expected from general
physical arguments. However, before making serious comparisons with
experiments, it should be kept in mind that unlike the nodal
quasiparticles, the off-nodal quasiparticles are in general
sensitive to the fine tuning of Hamiltonian parameters. In
particular, we find the anisotropy of the quasiparticle weight on
the Fermi surface depends crucially on the value of $t^{''}$. We
thus can not exclude the possibility that some other more generic
mechanism is responsible for the observed nodal-antinodal dichotomy.

Finally, we note that the backflow effect will also cause
renormalization of the ground state. In our calculation, we have
assumed implicitly that such a renormalization is not strong enough
to induce Fermi surface reconstruction, in which case our
calculation would be totally invalid. In the absence of the Fermi
surface reconstruction, such a renormalization effect on the ground
state can in principle be taken into account in our theory to arrive
at a more consistent description of the quasiparticle properties.
When the backflow effect is strong enough to cause Fermi surface
reconstruction, there arises the interesting and exotic possibility
of forming a small Fermi pocket without any symmetry breaking. An
important problem then is whether the small Fermi pocket has a
quantized volume. We leave these issues to future study.

Tao Li is supported by NSFC Grant No. 10774187 and National Basic
Research Program of China No. 2007CB925001, Fan Yang is grateful for
the NSFC Grant No.10704008.

\section{Appdenix A}
In this appendix, we show the rank of the basis functions
$|\mathrm{k},\mathrm{q},\sigma\rangle$ is lower than that of the
mean field states
$\gamma^{\dagger}_{\mathrm{k}-\mathrm{q},\sigma}b^{\dagger}_{\mathrm{q}}|\mathrm{SBMF}\rangle$
by one. This can be shown by proving the following identity
\begin{equation}
\sum_{\mathrm{q}}u_{\mathrm{k}-\mathrm{q}}|\mathrm{k},\mathrm{q},\sigma\rangle=0,
\end{equation}
in which
$u_{\mathrm{k}}=\frac{1}{2}(1-\frac{\xi_{\mathrm{k}}}{E_{\mathrm{k}}})$.

Using the Bogliubov transformation
$f^{\dagger}_{\mathrm{k}\sigma}=u_{\mathrm{k}}\gamma^{\dagger}_{\mathrm{k}\sigma}+v_{\mathrm{k}}\gamma_{-\mathrm{k},\bar{\sigma}}$,
we have
$u_{\mathrm{k}-\mathrm{q}}\gamma^{\dagger}_{\mathrm{k}-\mathrm{q},\sigma}=f^{\dagger}_{\mathrm{k}-\mathrm{q},\sigma}-v_{\mathrm{k}-\mathrm{q}}\gamma_{-\mathrm{k}+\mathrm{q},\bar{\sigma}}$.
As
$\gamma_{-\mathrm{k}+\mathrm{q},\bar{\sigma}}|f-\mathrm{BCS}\rangle=0$,
we have
\begin{eqnarray}
&&\sum_{\mathrm{q}}u_{\mathrm{k}-\mathrm{q}}|\mathrm{k},\mathrm{q},\sigma\rangle\nonumber\\
&=&\sum_{\mathrm{q}}\mathrm{P}_{N_{h}+1}\mathrm{P}_{\mathrm{s}}f^{\dagger}_{\mathrm{k}-\mathrm{q},\sigma}b^{\dagger}_{\mathrm{q}}|\mathrm{SBMF}\rangle\nonumber\\
&=&\sum_{i}\mathrm{P}_{N_{h}+1}\mathrm{P}_{\mathrm{s}}f^{\dagger}_{i,\sigma}b^{\dagger}_{i}|\mathrm{SBMF}\rangle\nonumber\\
&=&0,
\end{eqnarray}
in which the no double occupancy constraint has been used in the
final step.

\section{Appdenix B}
In this appendix, we derive the expression for the electron spectral
function in the approximation that $|\mathrm{k,q},\sigma\rangle$
form a complete set for the description of the quasiparticle
excitation and show that the spectral function so calculated does
satisfy the sum rule.

A bare hole created in the ground state is given by
$|\mathrm{k},\sigma\rangle_{0}=c_{\mathrm{k},\sigma}\mathrm{P}_{N_{e}}\mathrm{P}_{\mathrm{G}}|\mathrm{BCS}\rangle$.
Written in terms of the slave particles, it reads
\begin{eqnarray}
|\mathrm{k},\sigma\rangle_{0}=\sum_{\mathrm{q}}\mathrm{P}_{N_{h}+1}\mathrm{P}_{\mathrm{s}}f_{\mathrm{k-q},\sigma}b^{\dagger}_{\mathrm{q}}|\mathrm{SBMF}\rangle=\sum_{\mathrm{q}}\phi^{0}_{\mathrm{q}}|\mathrm{k,q},\sigma\rangle,
\end{eqnarray}
in which $\phi^{0}_{\mathrm{q}}=v_{\mathrm{k-q}}$. In the derivation
we have used the fact that $\mathrm{P}_{\mathrm{s}}$ and
$f_{i,\sigma}b^{\dagger}_{i}$ commute with each other.

Assuming that $|\mathrm{k,q},\sigma\rangle$ form a complete set, the
electronic spectral function can then be calculated as
\begin{equation}
A(\mathrm{k},\omega)=\sum_{n}|_{0}\langle\mathrm{k},\sigma|\mathrm{k},\sigma\rangle_{n}|^{2}\delta(\omega-(\mathrm{E}_{\mathrm{k},n}-\mathrm{E}_{g})),
\end{equation}
in which $|\mathrm{k},\sigma\rangle_{n}$ denotes the $n-$th
eigenvector of Eq.(15) and $\mathrm{E}_{\mathrm{k},n}$ is
corresponding eigenvalue. $\mathrm{E}_{g}$ is the variational ground
state energy. We thus have
\begin{equation}
A(\mathrm{k},\omega)=\sum_{n,\mathrm{q,q'}}|\phi^{0*}_{\mathrm{q}}\mathrm{O}_{\mathrm{q,q'}}\phi^{n}_{\mathrm{q'}}|^{2}\delta(\omega-(\mathrm{E}_{\mathrm{k},n}-\mathrm{E}_{g})),
\end{equation}
in which $\phi^{n}_{\mathrm{q}}$ is the $n-$th eigenvector of
Eq.(15) and satisfies the following orthonormal condition
\begin{equation}
\sum_{\mathrm{q,q'}}\phi^{n*}_{\mathrm{q}}\mathrm{O}_{\mathrm{q,q'}}\phi^{m}_{\mathrm{q'}}=\delta_{n,m}.
\end{equation}

The electronic spectral function so obtained satisfy the sum rule
$\int d \omega
A(\mathrm{k},\omega)=\langle\Psi|c^{\dagger}_{\mathrm{k},\sigma}c_{\mathrm{k},\sigma}|\Psi\rangle$.
In fact, from Eq.(5) we have
\begin{eqnarray}
\int d \omega
A(\mathrm{k},\omega)&=&\sum_{n}|_{0}\langle\mathrm{k},\sigma|\mathrm{k},\sigma\rangle_{n}|^{2}=_{0}\langle\mathrm{k},\sigma|\mathrm{k},\sigma\rangle_{0}\nonumber\\
&=&\langle\Psi|c^{\dagger}_{\mathrm{k},\sigma}c_{\mathrm{k},\sigma}|\Psi\rangle,
\end{eqnarray}
where we have used the fact that $|\mathrm{k},\sigma\rangle_{n}$
forms a complete set in the space spanned by
$|\mathrm{k,q},\sigma\rangle$.

\end{document}